\documentclass[12pt]{article}
\usepackage{a4wide}
\usepackage{amssymb, amsmath}
\usepackage[normalem]{ulem}
\usepackage[utf8]{inputenc}
\usepackage[colorlinks]{hyperref}
\usepackage[usenames,dvipsnames]{color}
\hypersetup{
     breaklinks=true,
    pdfstartview={FitH},    % fits the width of the page to the window
    colorlinks=true,       % false: boxed links; true: colored links
    linkcolor=blue,          % color of internal links
    citecolor=red,        % color of links to bibliography
    filecolor=magenta,      % color of file links
    urlcolor=blue,           % color of external links
    anchorcolor=green,      % Color for anchor text
    linktocpage=true
}

\usepackage[style=nature, citestyle=numeric-comp, sorting=none, backend=biber, maxnames = 5]{biblatex}
\bibliography{References.bib}

\newcommand{\bea}{\begin{eqnarray}}
\newcommand{\eea}{\end{eqnarray}}
\renewcommand{\d}{{\mathrm{d}}}
\renewcommand{\[}{\left[}

\renewcommand{\)}{\right)}

\def\be{\begin{equation}}
\def\ee{\end{equation}}

\begin{document}

{\renewcommand{\thefootnote}{\fnsymbol{footnote}}
		
\begin{center}
{\LARGE Is the CMB revealing signs of pre-inflationary physics?} 
\vspace{1.5em}

Suddhasattwa Brahma$^{1}$\footnote{e-mail address: {\tt suddhasattwa.brahma@gmail.com}} and 
Jaime Calder\'on-Figueroa$^{2}$\footnote{e-mail address: {\tt jrc43@sussex.ac.uk}}

\vspace{1.5em}

$^{1}$Higgs Centre for Theoretical Physics, School of Physics and Astronomy,\\ University of Edinburgh, Edinburgh EH9 3FD, UK\\[2mm]

$^2$Astronomy Centre, University of Sussex, Falmer, Brighton, BN1 9QH, UK\\[2mm]

\vspace{1.5em}
\end{center}
}

\setcounter{footnote}{0}

\begin{abstract}
\noindent Given the latest observational constraints coming from the joint analyses of the Atacama Cosmology Telescope, the Planck Satellite and other missions, we point out the possibility of reconciling fundamental particle-physics models of inflation with data by considering non--Bunch-Davies initial conditions for primordial density perturbations. 
\end{abstract}

On large scales, the Cosmological Principle states that the Universe is homogeneous and isotropic such that no point in it is `special'. Nevertheless, to explain the observed large scale structure of our cosmos, one needs to have initial conditions that are slightly inhomogeneous and anisotropic. It is well-known that having very tiny fluctuations (one part in $10^5$ on the last scattering surface) is sufficient to source the late time inhomogeneities that result in temperature anisotropies in the cosmic microwave background (CMB) or formation of galaxy clusters. The greatest achievement of inflation is to show that vacuum fluctuations, which must be present in any quantum theory, is sufficient to produce such initial conditions which are then stretched across cosmological distances by a phase of exponential expansion \cite{Pert2,Mukhanov:1982nu}. Since these fluctuations are amplified by gravitational instabilities, they were even smaller during inflation and hence one can trust linear perturbation theory to compute their statistics in our universe.

So what describes the microphysics of such a rapid phase of expansion? The canonical paradigm of single-field inflation posits that structures in the universe are sourced mainly by self-interactions of a scalar field that rolls very slowly down its own potential. One would then just have to adjust parameters in the potential so as to fit observations. However, this has turned out to be much more complicated than what was initially envisioned. Firstly, the density perturbations have thus far not revealed any deviations from Gaussianity, thereby creating a huge degeneracy in the parameter space of inflationary models. Moreover, we have not found any evidence for primordial gravitational waves yet which strongly constrains the energy scale of inflation. Finally, the power spectrum is almost scale-invariant with slightly more power on large scales. All of these observations taken together point towards a refinement of the viable fundamental models of inflation. 

The last data release from the \textit{Atacama Cosmology Telescope} (ACT) \cite{ACT:2025tim,ACT:2025fju}, combined with the observations of the Planck satellite \cite{Planck:2018jri,Planck:2018vyg} and the DESI DR1 dataset \cite{DESI:2024mwx, DESI:2024uvr}, have put stronger constraints on the spectral tilt $n_s$ such that it prefers a very flat spectrum, with a reported value $n_s = 0.9743 \pm 0.0034$ \cite{ACT:2025fju}. Current upper bounds on the so-called tensor-to-scalar ratio ($r$) put it at $r<0.036$  \cite{BICEP:2021xfz}. Since this directly constrains the energy scale of inflation in standard single-field models, it has ruled out large swaths of model space.

The central message of this note is to point out that we are at a crossroads in inflationary model-building and there are two different ways to reconcile with the latest CMB data, as revealed by P-ACT-LB-BK18 (ACT data combined with Planck, DESI DR1 and BICEP/\textit{Keck}). One path would be to tinker with well-motivated potentials for single-field inflation so that they fit data better. Take, for instance, Starobinsky inflation \cite{Inf5} (as a typical example of so-called $\alpha$-attractors \cite{Kallosh:2013yoa}) which is now disfavoured at 2$\sigma$ if this data holds. One can modify this model further with different forms of non-minimal couplings, or adding further symmetries, so as to make them consistent with observations \cite{Kallosh:2025rni, Aoki:2025wld}. However, such fine-tuning of models would then have to be done perpetually to fit future data. More generally, the preferred models of (minimally coupled) chaotic inflation with a $m^2 \phi^2$ or a $\lambda \phi^4$ potential have, by now, been ruled out by observations and one necessarily needs such additional couplings to make them viable again. This game would then need to be played again if, for instance, we find $r< 10^{-3}$ from future cosmological observations (LiteBird \cite{LiteBIRD:2022cnt}, Simons Observatory \cite{SimonsObservatory:2018koc}, and others.). 

Our main point is neither to criticize any of the above-mentioned models nor to say that there are currently no models of single-field inflation that fit the P-ACT-LB-BK18 data. Certainly, one can add further couplings, and fine-tune the additional coupling constants, to make viable models of $\alpha$-attractors \cite{Kallosh:2013tua} or non-minimal Higgs inflation \cite{Bezrukov:2007ep, Rubio:2018ogq}. However, in the absence of any overarching guiding principle, this would result in a fine-tuning problem that follows every new observation of the CMB or LSS data. The bigger problem is that constructing potentials, that are not protected by any symmetries, make them vulnerable to Planck-suppressed operators (see \cite{Baumann:2014nda, McAllister:2007bg} for a general discussion about this for stringy models). And thus, in the absence of a reliable EFT description of such phenomenological model-building, these models do not find natural UV-completions \cite{Bedroya:2019tba, Palti:2019pca}.

However, there is another way to interpret these results. What if we allow the initial state of scalar perturbations to be slightly modified from their Bunch-Davies (BD) values while sticking with potentials that are reliable from an EFT point of view? Inflation has been shown to be necessarily past-incomplete \cite{Borde:2001nh}, so it makes perfect sense to assume that there was a quantum gravity era preceding it which leaves its imprints in the initial state for inflation. More generally, since inflation begins at some finite time in the past, UV physics could signal a deviation from the standard Bunch-Davies initial conditions. Such a general initial state does not even need to be Gaussian or pure \cite{Agarwal:2012mq}, but we will only restrict the discussion to Bogoliubov rotations of the Bunch-Davies modes for this note.  Various deviations from standard single-field slow roll dynamics can lead to such a state, such as due to a previous radiation-dominated era \cite{Vilenkin:1982wt}, due to multi-field dynamics \cite{Shiu:2011qw}, having a non-attractor phase \cite{Lello:2013mfa}, resulting from a false vacuum \cite{Sugimura:2013cra}, due to a phase of anisotropic expansion \cite{Dey:2011mj,Dey:2012qp}, or due to some UV physics \cite{TP1, Kempf:2000ac,Kaloper:2002uj, Hui:2001ce, Schalm:2004qk, Ashoorioon:2004wd, Agullo:2015aba}.  

The only question that remains to be answered is if there has been sufficient e-foldings of inflation to wipe away the signatures of such pre-inflationary dynamics. Since the Bunch-Davies vacuum acts as a quantum attractor, there is an analogue of the no-hair theorem for de Sitter expansion which says that a small non--Bunch-Davies (NBD) component would be wiped away given enough expansion. And, for adiabatic initial conditions, such trans-Planckian effects do not affect late-time observations \cite{Burgess:2020nec}. Instead of going into theoretical prejudices of whether, and under which conditions, one should expect the imprint of trans-Planckian modes in the CMB data \cite{Collins:2005nu}, we look into this from a phenomenological point of view. We argue that the current constraints due to P-ACT-LB-BK18 indicate a slight preference for such a non--Bunch-Davies initial state if we want to stick to principled, well-motivated models of single-field inflation.

Expressing the non--Bunch-Davies initial modes as \cite{Ganc:2011dy, Brahma:2013rua}
\begin{eqnarray}
v_k(\eta) = \alpha_k u_k^{\rm BD}(\eta) + \beta_k u^{\rm BD}_k(\eta)\,,
\end{eqnarray}
where $u^{\rm BD}_k$ denotes the standard Bunch-Davies mode functions. Imposing the constraint $|\alpha_k|^2 - |\beta_k|^2 = 1$, one can rewrite the correction to the dimensionless scalar power spectrum for super-horizon modes as 
\begin{eqnarray}
\mathcal{P}_\zeta := P_\zeta^{\rm BD} \gamma_k = P_\zeta^{\rm BD}
\left[1 + 2 N_k + 2 \sqrt{N_k \left(1+N_k\right)} \cos\Theta_k\right]\,,\label{NBD_PS_Scal}
\end{eqnarray}
where $N_k = |\beta_k|^2$ stands for the expectation value of the number of particles in the excited state and $\Theta_k$ is the relative phase difference between $\alpha_k$ and $\beta_k$. Given this modification to the power spectrum, we can write the modification to the spectral tilt as
\begin{eqnarray}
  n_s -1 = (n_s-1)^{\rm BD} +  \left[ \frac{\d \ln \gamma_k}{\d\ln k}\right]\,,
\end{eqnarray}
and the correction to running of the tilt as 
\begin{eqnarray}
  \frac{\d n_s}{\d \ln k} = \left(\frac{\d n_s}{\d \ln k}\right)^{\rm BD} + \left[ \frac{\d^2 \ln \gamma_k}{(\d \ln k)^2}\right]\,.
\end{eqnarray}

We can choose a parametrisation
\begin{eqnarray}
    \beta_k \sim \begin{cases}
    0\,, & \text{when $k>k_{\rm cut}$}\\
    f(k), & \text{when $k<k_{\rm cut}$}
  \end{cases} 
\end{eqnarray}
where $f(k)$ has a mild $k$-dependence and $k_{\rm cut}$ denotes a cutoff above which the modes must be in their Bunch-Davies vacuum to satisfy the Hadamard condition. A crude physical model that can enforce such a choice is given by \cite{Holman:2007na}
\begin{eqnarray}\label{eq:param}
N_k = N_k^{0} \ e^{-k^2/\(M a(\eta_0)\)^2}\,,    
\end{eqnarray}
where the physical cutoff $M$ is the scale of new physics which must be bigger than the Hubble parameter during inflation for a consistent EFT description. One must assume that all modes of observational interest must be below this cutoff scale at $\eta_0$, the beginning of inflation, so that there is no step in $P_\zeta$. One cannot have a large value for $|f(k)|^2$ so as to avoid having a large $f_{\rm NL}$ for the local shape in the scalar bispectrum \cite{Holman:2007na, Ganc:2011dy, Agullo:2010ws, Flauger:2013hra} since this is tightly constrained by observations \cite{Planck:2019kim,Jung:2025nss}. Moreover, the standard backreaction constraints must be imposed on $\beta_k$ so that background slow-roll trajectory is not upset by the energy density of this excited state \cite{Holman:2007na}.

Instead of a systematic study of the entire parameter of non--Bunch-Davies initial states (which would necessarily also have to include mixed initial density matrices going beyond pure, Gaussian states \cite{Agarwal:2012mq}), we shall now give an example of how corrections from having even a tiny NBD component (parametrised by a small $N_k$) can still change the spectral tilt by a sufficient amount to make particle-physics models of inflation consistent with P-ACT-LB. More specifically, we consider Starobinsky \cite{Inf5}, or its closely-related version of Higgs inflation \cite{Rubio:2018ogq}, and show that these models can be consistent with current data by allowing for relatively moderate excited states. And any amount of amplification to $P_\zeta$ naturally further modifies $r$, and this can be used to further constrain $N_k$ from future observations of $r$.

Let us assume for the sake of simplicity that $\theta_k = \theta$ $\forall\ k$. Then, 
\begin{align}
    \frac{\d \ln \gamma_k}{\d \ln k } & = \frac{(1+2N_k) \cos \theta + 2\sqrt{N_k (N_k + 1)} }{\gamma_k \sqrt{N_k (N_k + 1)}} \frac{\d N_k}{\d \ln k} = \Delta n_s\;, 
\end{align}
where $d N_k/\d \ln k = -2 N_k (k/M a(\eta_0))^2$ for the parametrisation shown in Eq.~\eqref{eq:param}. Clearly, the rate of change is suppressed for short and long wavelengths relative to the cutoff scale. The choice of this scale is flexible, provided the aforementioned (backreaction) constraints are met. As a rule of thumb, a small value of $N_k^0$ ensures consistency.  

To illustrate this with an easy example, consider the case of Starobinsky inflation, with $V(\phi) = V_0 (1-\exp(-\sqrt{2/3}\phi))^2$. This model predicts a spectral index $n_s = 0.962 $ for $N = 50$ e-folds, and $n_s = 0.968$ for $N = 60$, which is in slight tension with the expected value of $0.974$. Unsurprisingly, this can be changed by considering NBD initial states at minimal cost. 
For $N = 60$, a correction of $\Delta n_s \simeq 0.006$ is required. Choosing $a(\eta_0)M \simeq 5 k_*$, the pivot scale is sufficiently inside the horizon at the desired scale/initial time $\eta_0$ (roughly 7 e-folds before crossing for $M \simeq 10 V^{1/4}$, but it reduces to 3 e-folds before horizon crossing for $M = 100 H$). Setting $\theta = \pi$, agreement with observations is achieved for $N_k^0 = 5.89\times 10^{-3}$, corresponding to $\gamma_{k_*} \simeq 0.86$. As a result, the power spectrum is suppressed relative to the Bunch--Davies case, and the tensor-to-scalar ratio is enhanced, yielding $r = 16\epsilon/\gamma_{k_*} \simeq 0.0034$. 
For $N = 50$, a larger correction is necessary, with $\Delta n_s = 0.012$. This, in turn requires a larger but still small $N_k^0 \simeq 2.4 \times 10^{-2}$ for $\theta = \pi$. This yields $\gamma_{k_*} \simeq 0.74$, and $r \simeq 0.0057$. 

The naive expression we employed for $N_k$ successfully restored the Starobinsky model's consistency with the observational constraints on $n_s$. However, this approach leads to a deviation beyond the reported constraint for the running of the spectral index, $\alpha_s = 0.0062 \pm 0.0052$. The Starobinsky model predicts $\alpha_s \simeq -0.00067$ for $N = 50$ and $\alpha_s \simeq -0.00048$ for $N = 60$, which for the model above turns into $\alpha_s = 0.023$ and $0.011$, respectively, once NBD are considered as before. Although these values are not ruled out, we can do a better fitting of the model with current constraints by using the following parametrisation for $\gamma_k$:  
\begin{align}
    \gamma_k = c_1 + \frac{1-c_1}{1 + (e k_*/k)^{c_2}}\;,
\end{align}
which is a sigmoid-like function. This ensures that $\gamma_k \to 1$ ($N_k \to 0$) for large $k$, while for small $k$, it asymptotically approaches $\gamma_k \sim c_1$. The parameter $c_2$ determines the sharpness of the transition between these two regimes. Importantly, since $c_1 < 1$, $N_k$ remains well-controlled.  
For $N = 50$, consistency with the expected $\alpha_s$ requires $\{c_1, c_2\} = \{0.945, 1.137\}$, corresponding to $\gamma_{k_*} \simeq 0.96$. Similarly, for $N = 60$, we find $\{c_1, c_2\} = \{0.974, 1.651\}$, yielding $\gamma_{k_*} \simeq 0.978$.  However, we emphasise that we only use the Starobinsky model as a representative example to illustrate our main message, and one can employ such general initial states to any model of inflation that one prefers to see if such modifications can make the model viable with observations \cite{Ashoorioon:2014nta}. We have achieved this for the studied case without significant difficulty by using generic parametrisations that readily satisfy the self-consistency relations imposed for excited initial states. Also, note that there is a large parameter space of such states if one allows $\theta_k$ to vary or choose other values for it, as well as if one allows the tensor modes to also be in a NBD state \cite{Brahma:2019unn}. However, the latter is difficult to justify from a microscopic model.

The main point which we are making in this paper is that observations seem to disfavour the simplest models of single-field inflation which can be derived from some fundamental principle and are robust against UV-corrections in the usual sense of having a controlled EFT description for them. While it is indeed possible to introduce new couplings to make some of these models viable, this seems less satisfactory unless there is a symmetry principle or UV-theory to guide us. What we suggest instead is to consider the possibility that inflation does indeed proceed according to some theoretically well-motivated action; however, the initial conditions are not those of the Bunch-Davies vacuum state. This physically makes sense since inflation inevitably starts at some finite time and traces of pre-inflationary dynamics, either those of an excited heavy field or those of trans-Planckian physics where gravity is strong, get imprinted in the short distance behaviour of the initial state for fluctuations. 

We do not claim that doing this is any less fine-tuning than playing with monomial potentials as is often done. However, this is just as good an explanation and we explore the possibility of interpreting current observations as telling us that the relics of such trans-Planckian physics do not, in fact, get washed away by inflation and they can be used to constrain the initial state of fluctuations. What would be interesting would be to systematically derive such an initial state from some UV-theory and see if the generic non--Bunch-Davies modes imply modifications which better fit observations. However, it is important to emphasise that such a possibility still exists where we might be able to get a glimpse of pre-inflationary dynamics even when assuming inflation as the paradigm for the early universe.

\section*{Acknowledgments}
SB is supported in part by the Higgs Fellowship and by the STFC Consolidated Grant ``Particle Physics at the Higgs Centre''. JCF is funded by the STFC under grant number ST/X001040/1.

\printbibliography

\end{document}